\documentclass[journal,onecolumn,draftclsnofoot,10pt]{IEEEtran}

\usepackage{times}
\usepackage{epsfig}
\usepackage{amsmath}
\usepackage{amsfonts}
\usepackage{graphicx}
\usepackage{amssymb}
\usepackage{amstext}
\usepackage{latexsym}
\usepackage{color}
\usepackage{ifthen}
\usepackage{multirow}
\usepackage{verbatim}
\usepackage{array,tabularx}
\usepackage[mathscr]{euscript}
\usepackage{tikz}
\usepackage{units}

\newcommand{\be}{\begin{equation}}
\newcommand{\ee}{\end{equation}}
\newcommand{\bear}{\begin{eqnarray}}
\newcommand{\eear}{\end{eqnarray}}
\newcommand{\bears}{\begin{eqnarray*}}
\newcommand{\eears}{\end{eqnarray*}}
\newcommand{\bi}{\begin{itemize}}
\newcommand{\ei}{\end{itemize}}
\newcommand{\ben}{\begin{enumerate}}
\newcommand{\een}{\end{enumerate}}

\newtheorem{theorem}{Theorem}

\IEEEoverridecommandlockouts

\begin{document}
\title{New Codes and Inner Bounds for Exact Repair in Distributed Storage Systems}

%


\author{
\IEEEauthorblockN{Sreechakra Goparaju  \hspace{2cm}Salim El Rouayheb \hspace{2cm}Robert Calderbank}
\thanks{S. Goparaju is with the EE Department, Princeton University, USA (e-mail: goparaju@princeton.edu)} 
\thanks{S.  El Rouayheb is with the ECE Department, Illinois Institute of Technology,  USA (e-mail: salim@iit.edu).}
\thanks{R.  Calderbank is with the EE Department of Electrical Engineering, Duke University, USA (e-mail: robert.calderbank@duke.edu).}
}

\maketitle

\begin{abstract}
We study the exact-repair tradeoff between storage and repair bandwidth in distributed storage systems (DSS). We give new inner bounds for the tradeoff region and provide code constructions that achieve these bounds.

 \end{abstract}

\section{Introduction}\label{sec:intro}
The study of erasure codes which repair single node failures in a distributed storage system more efficiently was initiated by Dimakis et al in \cite{DGWWR10}. The codes, called {\em regenerating codes}, are constructed on a system of $n$ storage nodes and store a file of size ${\cal M}$, such that the data stored in any $k$ nodes is sufficient to recover the ${\cal M}$ symbols. Furthermore, each node stores $\alpha$ symbols and each failed node requires a transmission of $\gamma$ symbols, termed as the {\em repair bandwidth}, by $d$ other nodes, called {\em helper} nodes ($d$ is also known as the {\em repair degree}). An optimal tradeoff between storage and repair-bandwidth ($\alpha$ and $\gamma$) is characterized in \cite{DGWWR10} for the so-called case of {\em functional repair}, where the repaired (or replaced) node need not store the exact copy of the data present in the failed node. This tradeoff region is given by the following expression:
\begin{eqnarray}\label{eq:fctrepair}
\mathcal{M}&\le& \sum_{i=0}^{k-1}\min\left\{\alpha,(d-i)\frac{\gamma}{d}\right\}.
\end{eqnarray}

However, when an exact duplication of data onto the repaired node, or {\em exact repair}, is required, this characterization remains an open problem in general. Since exact repair is a stricter constraint than functional repair, the tradeoff curve in (\ref{eq:fctrepair}) forms an outer bound to the exact-repair region for $\alpha$ and $\gamma$ (for a given ${\cal M},n,k$ and $d$). The extremal points of this tradeoff have attracted the most attention, being the optimal points in terms of $\alpha$ and $\gamma$. These are respectively called the {\em minimum storage regenerating} (MSR) point, with
\begin{eqnarray}\label{eq:MSR}
\begin{array}{ccccl}
\alpha &=& \alpha_{{\sf msr}} &=& \dfrac{{\cal M}}{k}, \,\,\,\textrm{and}\\[0.3cm]
\gamma &=& \gamma_{{\sf msr}} &=& \dfrac{d{\cal M}}{k(d-k+1)},
\end{array}
\end{eqnarray}
and the {\em minimum bandwidth regenerating} (MBR) point, with
\begin{eqnarray}
\begin{array}{ccccl}
\alpha &=& \alpha_{{\sf mbr}} &=& \dfrac{2d{\cal M}}{k(2d-k+1)},\,\,\,\textrm{and}\\[0.3cm]
\gamma &=& \gamma_{{\sf mbr}} &=& \dfrac{2d{\cal M}}{k(2d-k+1)},
\end{array}
\end{eqnarray}
Both points have been shown to be achieveable for exact repair for all $(n,k,d)$, using finite-length or asymptotic codes; see, for example, \cite{RSK11, PDC11, CJMRS13}, and \cite{WTB11}.

\begin{figure}[t]
\begin{center}
\includegraphics[trim =12mm 0mm 0mm 0mm, clip,width=0.44\textwidth]{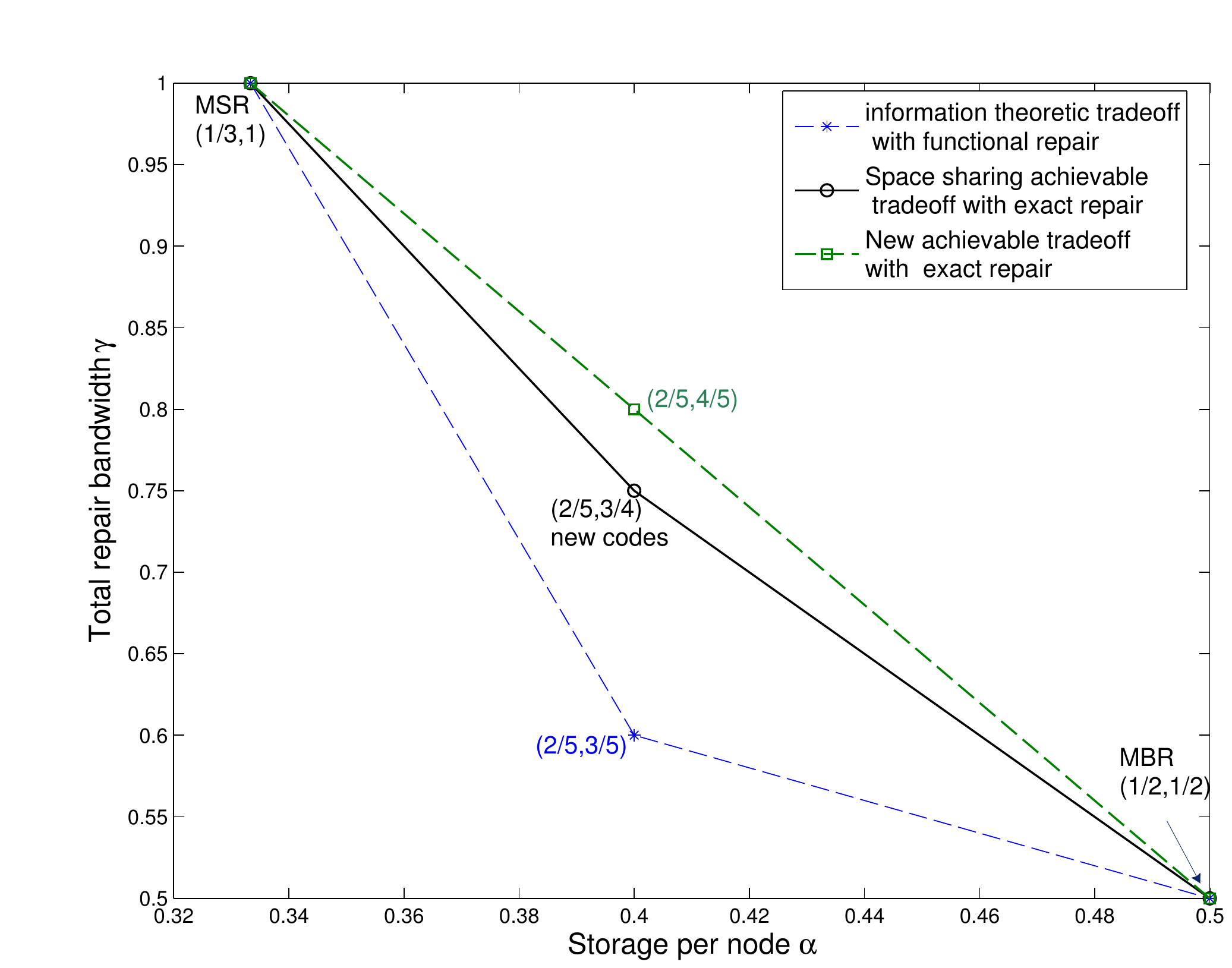}
\caption{ Example of the new achievable exact-repair tradeoff  for the $(5,3,3)$ case for a file size $\mathcal{M}=1$. The code construction achieving the  point $(\frac{2}{5},\frac{3}{4})$ is described in Sec.~\ref{sec:example}.  }
    \label{fig:tradeoff}
   \end{center}
\end{figure}

Recently, Tian \cite{T13} fully determined the optimal tradeoff for the case of an $(n,k,d)=(4,3,3)$ DSS. The optimal tradeoff regions for all other cases of $(n,k,d)$ are still under investigation. Indeed, for some time, no inner bounds existed that were tighter than the space-sharing (or time-sharing) bound between the MSR and the MBR points. However, recently, new codes have been independently discovered, which improve upon the space-sharing bound, and lie in the so-called {\em intermediate} region, that is, between the MSR and the MBR points. In \cite{TAV13}, block-designs and erasure codes were used in a layered structure to build intermediate codes which are simple to implement. In \cite{SK13}, layering is used to construct codes first for the parameters $(n,k,k)$, i.e., when $d=k$, and then extended to $(n,k,d)$, when $d > k$. In \cite{E13}, new $(n,k,d)$ regenerating codes at intermediate points are formed using MSR points for lower parameters.

{\em Contribution:} We construct new intermediate codes which in turn lead to newer inner bounds for the storage/repair bandwidth tradeoff region for exact repair. As an example, for a $(5,3,3)$ DSS, our codes achieve the new point $(\alpha,\gamma)=(2/5,3/4)$, which leads to the new achievable tradeoff shown in Fig. \ref{fig:tradeoff}. This example is explained in detail in Section \ref{sec:example}, and is used to provide an intuition behind our code construction. Our technique is inspired by the constructions in \cite{SK13} and \cite{E13}, and the study of heterogeneous DSS in \cite{ERHP13}. We use MSR code constructions for smaller codes to achieve intermediate points for larger parameters. We then generalize this technique in order to improve the repair bandwidth $\gamma$ while keeping the rest of the parameters -- $n,k,d,\alpha$, and ${\cal M}$ -- constant. In Section \ref{sec:d=k}, we describe our code construction and the ensuing tradeoff inner bound for the case when $d=k$. In Section \ref{sec:d>k}, we extend this construction to the case when $d>k$. We then describe a second construction for this regime which uses ideas in \cite{SK13} in a flavor similar to the way we improve the repair bandwidth.
We show that this construction proves as useful as the first construction by plotting the tradeoff region for an example DSS; see Fig. \ref{fig:61-55-59}.
Finally, we conclude in Section \ref{sec:conclusion}.

\section{Primer: A $(5,3,3)$ Example}\label{sec:example}
 
To illustrate our results, we start with a $(5,3,3)$ DSS consisting of five nodes $v_1,\dots,v_5$. We give a new regenerating code for this DSS, which leads to an improved inner (achievable) bound on the storage/repair bandwidth tradeoff curve for this DSS under exact repair. 
The general results and proofs will be detailed in the subsequent sections. 

\begin{figure}[t]
\begin{center}
\includegraphics[]{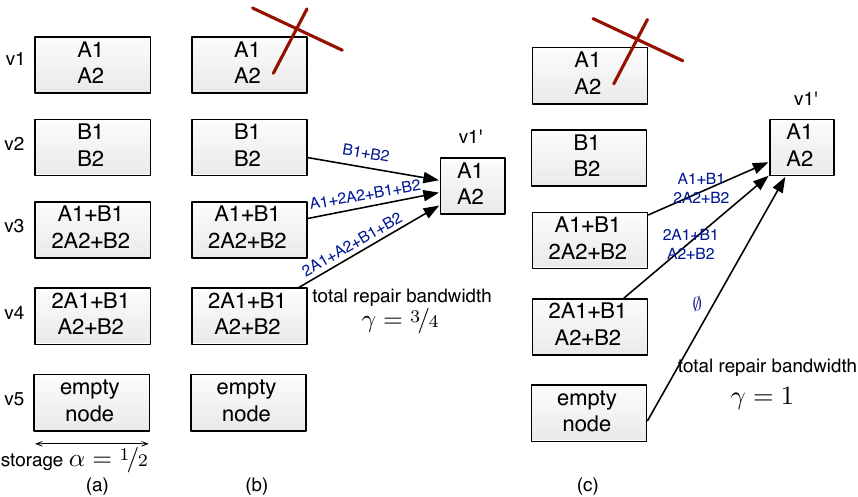}
\caption{Part (a) A $(5,3,3)$ code constructed by appending a $(4,2,3)$ MSR code \cite{SR10} to an empty node. Part (b) shows the repair of a non-empty node in the $(5,3,3)$ system when the $d=3$ helper nodes belong to the  MSR code. Part (c) shows the repair when the $d$ helper nodes consist of only $2$ nodes in the MSR code.
}
    \label{fig:ex}
   \end{center}
\end{figure}

The idea is to start with an MSR code for a smaller $(4,2,3)$ DSS with $(\hat{n},\hat{k})=(4,2)$ consisting of the nodes $\{v_1,\dots,v_4\}$ as in Fig.~\ref{fig:ex}(a). Here, we pick  the $(4,2,3)$ MSR code  from \cite{SR10}. Each node stores half a symbol ($\hat{\alpha}=1/2$), making the total stored file size ${\cal M}=1$. We extend this code by adding an empty node $v_5$. The resulting DSS is heterogeneous since different nodes store different amounts of data. Moreover, the total repair bandwidth depends on which helper nodes participate in the repair process. There are two cases here. 

{\em Case 1:} All the $d=3$ helper nodes belong to the small MSR code, i.e., are chosen from $\{v_1,\dots,v_4\}$. Here, we can achieve the optimal repair bandwidth given by the MSR code, i.e., the repair bandwidth is $3/4$. 

{\em Case 2:} Only two nodes of the small MSR code help in the repair, the third node being $v_5$. In this case, the repair proceeds by downloading the whole file from the small code (and nothing from node $v_5$). The repair bandwidth is $1$. 

From \cite{ERHP13}, we know that we can construct a homogeneous code by ``glueing" together all the  $n!=120$ permuted copies of the heterogeneous code. This glued code achieves the average storage per node $\alpha$ and average repair bandwidth $\gamma$ of the heterogeneous code. For the code in Fig.~\ref{fig:ex}, a fraction of $4/5$ of the nodes are non-empty. Furthermore, for the repair of a non-empty node, a fraction of $1/4$ of helper sets consists of three helper nodes ({\em Case 1}) in the $(4,2,3)$ code. Thus, we have
\begin{eqnarray*}
\begin{array}{ccccccc}
\alpha &=& \dfrac{2}{5}, &\textrm{and}& \gamma_i &=& \dfrac{15}{16}, \,\,\textrm{for}\,\,\, i \in \{1,2,3,4\},
\end{array}
\end{eqnarray*}
where $\gamma_i$ is the average repair bandwidth when repairing node $v_i$, $i=1,\dots,5$. The repair bandwidth $\gamma_5$ when node $v_5$ fails is zero since no data needs to be downloaded. Therefore, the average total repair bandwidth, over all failures and all choices of helper nodes, is $\gamma=3/4$. 

The tradeoff curve for functional repair \cite{DGWWR10} is given by the piecewise linear curve joining the MSR point $(\alpha,\gamma)=(1/3, 1)$, an intermediate point $(2/5,3/5)$ and the MBR point $(1/2,1/2)$  (see Fig.~\ref{fig:tradeoff}). The MSR and MBR points are also achievable with exact repair \cite{CJMRS13, RSK11}. By space sharing, we can achieve any point on the straight line joining the MSR and MBR points. The code constructions in \cite{E13} and \cite{SK13} both correspond to the point $(2/5,4/5)$  on the space sharing curve for the $(5,3,3)$ example. The proposed code lies beneath this curve in Fig.~\ref{fig:tradeoff} and thus leads to an improved achievable tradeoff curve for exact repair.

\section{Code Construction for $d=k$}\label{sec:d=k}

Our goal is to build $(n,k,d)$ regenerating codes with exact repair for intermediate points on the  tradeoff curve. We focus first on the case when $d=k$ and address the case when $d>k$ in the next section. We follow the method in \cite{E13} and build the $(n,k,d)$ code from a small $(\hat{n},\hat{k})$ code to which we append $n-\hat{n}$ empty nodes as depicted in Fig~\ref{fig:GenCons}. In the previous example, we had $(n,k,d)=(5,3,3)$ and $(\hat{n},\hat{k})=(4,2)$. Notice that we do not specify the repair degree $\hat{d}$ yet.

\subsection{Construction}
 The key ingredient in our construction is to pick the small code to be an MSR code with minimum repair bandwidth  for all possible values of the repair degree  $\hat{d}=\hat{k}, \hat{k}+1,\dots, \hat{n}-1$, simultaneously, that is, the repair bandwidth $\hat{\gamma}_{\hat{d}}$ when the repair degree is $\hat{d}$ is given by
 \begin{eqnarray}\label{eq:repairBandwidth}
  \hat{\gamma}_{\hat{d}}&=&\frac{\hat{d}}{\hat{d}-\hat{k}+1}\,\hat{\alpha},
  \end{eqnarray}
 where $\hat{\alpha}$ is the storage per node in the small code; see (\ref{eq:MSR}).
  Constructions of MSR codes that can asymptotically achieve the minimum repair bandwidth for any repair degree  were described in \cite{CJMRS13} and are based on the interference alignment method. This code property, referred to as {\em universality} in \cite{CJMRS13}, has been studied in depth in \cite{ATVC13}, where it is referred to as {\em opportunistic repair}. Note that for the special case of  $\hat{n}-1=\hat{k}+1$, as in the example, any MSR code has this property since for $\hat{d}=\hat{n}-1=\hat{k}+1$, optimal repair bandwidth follows from the definition of an MSR code, and for $\hat{d}=\hat{k}$, the optimal repair requires downloading the whole file.
 
Hence we do not need to worry about $\hat{d}$  when choosing the parameters  of the small code (as long as $\hat{n}-1\geq\hat{k}$, which is true for a non-trivial small code). Since file reconstruction and exact repair in the big code are inherited from the small code, we need to pick $\hat{k}$ to make sure that among the  $k$ nodes from $\{v_1,\dots, v_n\}$ contacted by a user, there are always $\hat{k}$ nodes from $\{v_1,\dots,v_{\hat{n}}\}$. This is achieved by choosing $\hat{n}$ and $\hat{k}$ such that the big code and small code have the same number of parity nodes, i.e., 
\begin{eqnarray}\label{eq:n-d=k}
n-k=\hat{n}-\hat{k}. 
\end{eqnarray}
Also, since $d=k$, the repair process is guaranteed to contact at least $\hat{k}$ nodes from $\{v_1,\dots,v_{\hat{n}}\}$, and thus can be inherited from the small code.  We vary  $\hat{k}$ to take all the  integer values between  $1$ and  $k$ while choosing $\hat{n}=\hat{k}+(n-k)$.  Each value of $\hat{k}$ then results in a distinct regenerating code.  

 \begin{figure}[t]
 \begin{center}
\includegraphics[]{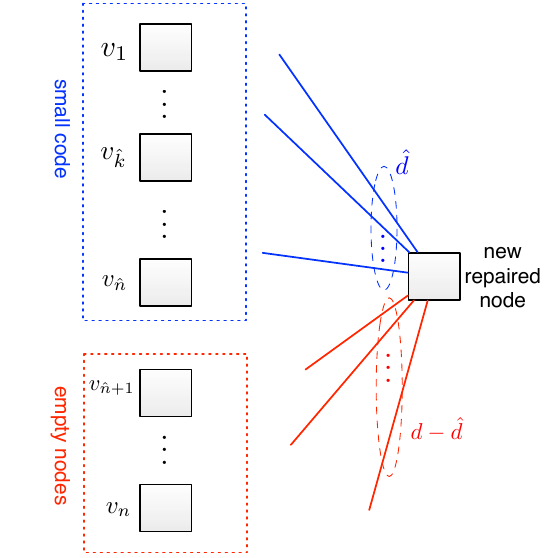}
\caption{ General code construction. }
    \label{fig:GenCons}
   \end{center}
\end{figure}
 
Ultimately, the code is formed of all the $n!$ permuted copies of the code in Fig.~\ref{fig:GenCons} glued together\footnote{It may  be possible to make the system homogeneous without using all the $n!$ permutations, but we don't worry about  this issue here.} as in \cite{ERHP13}. This code achieves the point $(\alpha,\gamma)$, where $\alpha$ and $\gamma$ are the average node storage capacity and repair bandwidth of the code in Fig.~\ref{fig:GenCons}, respectively. Since the storage per node in the small code is $\hat{\alpha}$, we obtain
 \begin{eqnarray}\label{eq:alpha-d=k}
\alpha&=&\frac{\hat{n}}{n}\,\hat{\alpha}.
 \end{eqnarray}
 
 Let $\gamma_1$ be the average repair bandwidth when repairing a non-empty node. Among all the possible sets of helper nodes,
 there is a fraction $P_{\hat{d}}$ (defined in (\ref{eq:gamma1-d=k})) of sets that consist of exactly $\hat{d}$ nodes in the small code, where $\hat{d}$ ranges from $\hat{k}$ to $\hat{d}_{{\sf max}}$ and $\hat{d}_{{\sf max}}=\min\{\hat{n}-1,d\}$. 
So we obtain
\begin{align}
\gamma_1 &\,\,=\,\, \sum_{\hat{d}=\hat{k}}^{\hat{d}_{{\sf max}}}\hat{\gamma}_{\hat{d}}P_{\hat{d}},\nonumber
\\
&\,\,=\,\, \sum_{\hat{d}=\hat{k}}^{\hat{d}_{{\sf max}}}\left(\frac{\hat{d}}{\hat{d}-\hat{k}+1}\right)\frac{\dbinom{\hat{n}-1}{\hat{d}}\dbinom{n-\hat{n}}{d-\hat{d}}}{\dbinom{n-1}{d}}\,\hat{\alpha}.\label{eq:gamma1-d=k}
\end{align}
Because the average repair bandwidth when repairing an empty node is zero, the overall average repair bandwidth $\gamma$ is given by
\begin{eqnarray}\label{eq:gamma-d=k}
\gamma &=& \sum_{\hat{d}=\hat{k}}^{\hat{d}_{{\sf max}}}\left(\frac{\hat{d}}{\hat{d}-\hat{k}+1}\right)\frac{\dbinom{\hat{n}-1}{\hat{d}}\dbinom{n-\hat{n}}{d-\hat{d}}}{\dbinom{n-1}{d}}\,\alpha.
\end{eqnarray}
  
  Since the small code is MSR,  the stored file can be of size up to 
  \begin{eqnarray}\label{eq:capacity-d=k}
  \mathcal{M}&=&\hat{k}\hat{\alpha}.
  \end{eqnarray}   
 
 \subsection{Inner Bound}
 From the above construction and from (\ref{eq:alpha-d=k}), (\ref{eq:gamma-d=k}), and (\ref{eq:capacity-d=k}), we have the following new achievable exact-repair tradeoff. Since it is possible that the points obtained by the construction do not enclose the MBR and the MSR points, we also add them to the tradeoff region.
   
\vspace{0.15cm} 
\begin{theorem}\label{th:innerbound-d=k}
There exist $(n,k,d)$ exact-repair regenerating codes, with $d=k$, that store a file of size $\mathcal{M}$ and achieve any point in the region $\sf{con}(\mathcal{R})$, the convex hull of $\mathcal{R}$, where 
\begin{align}
\mathcal{R}&= {\cal R}_1 \cup {\cal R}_{{\sf mbr}} \cup {\cal R}_{{\sf msr}},\\
\mathcal{R}_1&=\bigcup_{\hat{k}=1}^{k}\left\{ (\bar{\alpha},\bar{\gamma}): \bar{\alpha}\geq\alpha(\hat{k}),\bar{\gamma}\geq\gamma(\hat{k})\right\},\\
\mathcal{R}_{{\sf mbr}}&=\left\{ (\bar{\alpha},\bar{\gamma}): \bar{\alpha}\geq\alpha_{{\sf mbr}},\bar{\gamma}\geq\gamma_{{\sf mbr}}\right\}, \,\,\,\,\textrm{and}\\
\mathcal{R}_{{\sf msr}}&=\left\{ (\bar{\alpha},\bar{\gamma}): \bar{\alpha}\geq\alpha_{{\sf msr}},\bar{\gamma}\geq\gamma_{{\sf msr}}\right\}.
  \end{align} 
Here $\alpha(\hat{k})$ and $\gamma(\hat{k})$ are given by (\ref{eq:alpha-d=k}) and (\ref{eq:gamma-d=k}), respectively. Notice that $\hat{k}$ also determines $\hat{n}$ from (\ref{eq:n-d=k}), and $\hat{\alpha}$ is determined by (\ref{eq:capacity-d=k}).
\end{theorem}

\vspace{0.2cm}  
{\em Remark 1:} 
For the case of $d=k$, the two code constructions in \cite{E13, SK13} achieve the same points on the tradeoff, that are obtained when the summation in (\ref{eq:gamma-d=k}) contains only the term corresponding to $\hat{d}=\hat{k}$. 
The inner bound in Theorem~\ref{th:innerbound-d=k} is thus an improvement 
whenever the summation in \eqref{eq:gamma-d=k} has more than one term. This happens for all system parameters except when $k=d=n-1$, for which the three inner bounds coincide.

{\em Remark 2:} 
In general, we only know of asymptotic schemes \cite{CJMRS13} which achieve the repair bandwidth given in (\ref{eq:repairBandwidth}) for all feasible values of $\hat{d}$. However, for $n-k = 2$, the summation in (\ref{eq:gamma-d=k}) has summands corresponding to the two extreme cases of $\hat{d} = \hat{k}$ and $\hat{d} = \hat{k} + 1$, where the former requires each helper disk to transmit all its data as repair bandwidth, and the latter requires each helper disk to transmit only $1/(n-k)$ of its data. The latter is the optimal repair bandwidth for an $(\hat{n},\hat{k},\hat{n}-1)$ MSR code, for which several finite code constructions exist, e.g. \cite{RSK11,PDC11} and \cite{WTB11}.

%

\section{Code Constructions for $d > k$}\label{sec:d>k}
The construction in Section \ref{sec:d=k} can be generalized to the case when $d > k$ by using the same small code as before. The difference however arises during the repair of a failed node in the big code, as detailed in {\em Construction 1} below. 
A second construction, {\em Construction 2}, can be achieved using similar ideas, this time following the method in \cite{SK13}.

\subsection{Construction 1}\label{subsec:1}
As before, we construct the big $(n,k,d)$ regenerating code using the permuted copies of a small MSR $(\hat{n},\hat{k})$ code, where $\hat{n}-\hat{k}=n-k$, and which is repair-bandwidth-optimal for all possible degrees of repair $\hat{d} \in \{\hat{k},\hat{k}+1,\ldots,\hat{n}-1\}$.

Upon failure of a non-empty node, say $v_1$ (see Fig. \ref{fig:GenCons}), in the $(n,k,d)$ regenerating code, a fraction $P_{\hat{d}}$ of helper sets consist of exactly $\hat{d}$ nodes in the small code.
The repair bandwidth $\gamma_1$ is then given by
\begin{align}
\gamma_1 &\,\,=\,\, \sum_{\hat{d}=\hat{d}_{{\sf min}}}^{\hat{d}_{{\sf max}}}\hat{\gamma}_{\hat{d}}P_{\hat{d}},\\
&\,\,=\,\, \sum_{\hat{d}=\hat{d}_{{\sf min}}}^{\hat{d}_{{\sf max}}}\left(\frac{\hat{d}}{\hat{d}-\hat{k}+1}\right)\frac{\dbinom{\hat{n}-1}{\hat{d}}\dbinom{n-\hat{n}}{d-\hat{d}}}{\dbinom{n-1}{d}}\,\hat{\alpha},
\end{align}
where $\hat{d}_{{\sf min}} = d-(n-\hat{n})$, and $\hat{d}_{{\sf max}} = \min\left\{\hat{n}-1,d\right\}$ as defined earlier. Notice that unlike when $d = k$, any $d$ (helper) nodes must contain at least $d - (n-\hat{n})$ nodes in the small code, where it can be verified that $\hat{d}_{{\sf min}} > \hat{k}$.

Since the repair bandwidth $\gamma_i$ for the failure of an empty node $v_i$ is $0$, the overall repair bandwidth $\gamma$ is given by
\begin{eqnarray}\label{eq:gamma-d>k}
\gamma &=& \sum_{\hat{d}=\hat{d}_{{\sf min}}}^{\hat{d}_{{\sf max}}}\left(\frac{\hat{d}}{\hat{d}-\hat{k}+1}\right)\frac{\dbinom{\hat{n}-1}{\hat{d}}\dbinom{n-\hat{n}}{d-\hat{d}}}{\dbinom{n-1}{d}}\,\alpha.
\end{eqnarray}
The stored file size, as before, is given by ${\cal M} = \hat{k}\hat{\alpha}$, and the average storage per node is given by $\alpha = (\hat{n}/n)\hat{\alpha}$.

\subsection{Construction 2}\label{subsec:2}
When $d > k$, an alternative code can be constructed by viewing the big $(n,k,d)$ regenerating code as an $(n,d,d)$ regenerating code as obtained by {\em Construction 1} (in Section \ref{sec:d=k}), and calculating the {\em amount of information}\footnote{This can be viewed in different forms -- rank of the code generating matrix punctured at the columns that correspond to the $(n-k)$ nodes that are not selected, as in \cite{SK13}, or as the entropy of the random variables corresponding to the $k$ selected nodes.} that any set of $k$ nodes contains. This quantity ${\cal M}_k$ is the same for any set of $k$ nodes from the symmetry in {\em Construction 1}. It can be further verified that ${\cal M}_k$ is equal to the average amount of information contained in all the possible sets of size $k$ (in the heterogeneous code). Let ${\cal M}_d$ $(\ge {\cal M}_k$), be the file size of the $(n,d,d)$ code obtained from the construction, and let ${\cal M}_k$ be the size of the file that we wish to store in the DSS. The underlying motivation is to enable the recovery of the file from any $k$ nodes (which as we defined contain ${\cal M}_k$ amount of information). One way to achieve this is to concatenate a maximum rank distance (MRD) code and the $(n,d,d)$ code. To do so, the MRD codeword is constructed using a linearized polynomial with coefficients as the ${\cal M}_k$ symbols of our file. The output codeword of this code can then be viewed as a set of vectors which are evaluations of this polynomial on ${\cal M}_d$ points in a specific field, which are then fed as the input message file for the $(n,d,d)$ regenerating code constructed using {\em Construction 1}. The details of this concatenation are skipped here but can be found, for example, in \cite{SK13}. 
We now calculate the overall file size ${\cal M}$ ($= {\cal M}_k$) achievable.

Suppose that the small code being used is an $(\hat{n},\hat{k})$ MSR code, where now $\hat{n}-\hat{k}=n-d$. Let this code have $\hat{\alpha}$ units of storage per node and be repair-bandwidth optimal for all possible degrees of repair $\hat{d} \in \{\hat{k},\hat{k}+1,\ldots,\hat{n}-1\}$. Consider any set ${\cal K}$ of $k$ nodes in the big code. We noticed in Section \ref{sec:d=k} when $d=k$ that ${\cal K}$ contains at least $\hat{k}$ nodes of the small code in all permutations of the small code and $(n-\hat{n})$ empty nodes. When $d > k$, we use the property that the amount of information (entropy or linear dimension for the case of linear codes) contained in a set of $\omega$ nodes in the $(\hat{n},\hat{k})$ MSR code is given by $H_{\omega} = \min\{\omega,\hat{k}\}\,\hat{\alpha}$.

Therefore, if $Q_{\omega}$ is the fraction of permutations for which $\omega$ (and only $\omega$) of the nodes in the small code occur in ${\cal K}$, the average amount of information ${\cal M}_{k}$ in ${\cal K}$ is given by
\begin{align}
{\cal M}_k &\,\,=\,\, \sum_{\omega=\omega_{{\sf min}}}^{\omega_{{\sf max}}}H_{\omega}Q_{\omega},\\
&\,\,=\,\, \sum_{\omega=\omega_{{\sf min}}}^{\omega_{{\sf max}}}\min\{\omega,\hat{k}\}\frac{\dbinom{\hat{n}}{\omega}\dbinom{n-\hat{n}}{k-\omega}}{\dbinom{n}{k}}\,\hat{\alpha},\label{eq:capacity-d>k}
\end{align}
where $\omega_{{\sf min}} = \max\{1,\hat{n}-(n-k)\}$ and $\omega_{{\sf max}} = \max\{k,\hat{n}\}$ give the limits determined by the possible overlap sizes of the small code and ${\cal K}$.

Observe that the repair process is not disrupted during this process and hence the average repair bandwidth remains the same as in (\ref{eq:gamma-d=k}).

\subsection{Inner Bound}
A natural question arises whether one of the constructions for $d > k$ is better than the other. We show using an example that this is not necessarily true.

\begin{figure}[t]
\begin{center}

\includegraphics[trim =12mm 0mm 0mm 0mm, clip,scale=0.44]{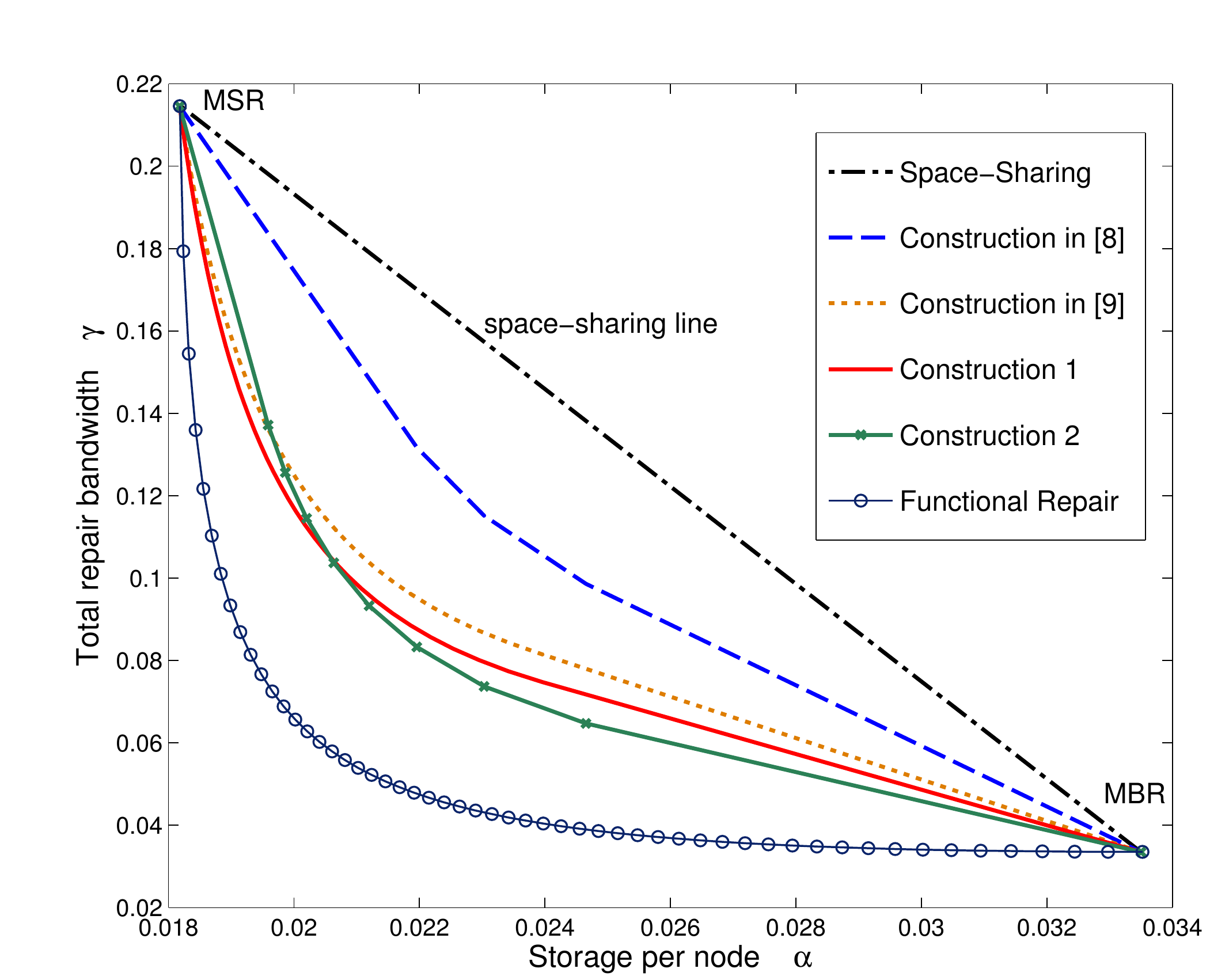}
\caption{Comparison among the different bounds existing in the literature and the new  inner bounds achieved by constructions 1 and 2 in Section~\ref{sec:d>k} for the case of $(n,k,d)=(61,55,59)$. The overall tradeoff region as defined in Theorem \ref{th:innerbound-d>k} is achieved by the convex hull of the two new constructions.
\vspace{-0.7cm}
}
    \label{fig:61-55-59}
\end{center}
\end{figure}

{\em Example:}
Consider\footnote{We considered this example for easy comparison with the adjacent case of $(61,55,60)$ considered in \cite{SK13}.} an $(n,k,d)=(61,55,59)$ DSS. Fig. \ref{fig:61-55-59} plots the inner bounds achieved for the storage/repair-bandwidth tradeoff using different code constructions. We see that for this example, the inner bound achieved by using the codes in \cite{E13} strictly encompasses the bound achieved by those in \cite{SK13}. However, {\em Construction 1} and {\em Construction 2}, do not exhibit such a relationship. Whereas {\em Construction 1} outperforms {\em Construction 2} at points closer to the MSR point, {\em Construction 2} generates a tighter inner bound near the MBR point.

It follows therefore that an overall inner bound must incorporate points from both constructions, as formalized in the next theorem.

\begin{theorem}\label{th:innerbound-d>k}
There exist $(n,k,d)$ exact-repair regenerating codes that store a file of size $\mathcal{M}=1$ and achieve any point in the region $\sf{con}(\mathcal{R})$, the convex hull of $\mathcal{R}$, where 
\begin{align}
\mathcal{R}&= {\cal R}_1 \cup {\cal R}_2 \cup {\cal R}_{{\sf mbr}} \cup {\cal R}_{{\sf msr}},\\
\mathcal{R}_1&=\bigcup_{\hat{k}_1=1}^{k}\left\{ (\bar{\alpha},\bar{\gamma}): \bar{\alpha}\geq\dfrac{\alpha(\hat{k}_1)}{{\cal M}(\hat{k}_1)},\bar{\gamma}\geq\dfrac{\gamma(\hat{k}_1)}{{\cal M}(\hat{k}_1)}\right\},\label{eq:regionConst1}\\
\mathcal{R}_2&=\bigcup_{\hat{k}_2=1}^{d}\left\{ (\bar{\alpha},\bar{\gamma}): \bar{\alpha}\geq\dfrac{\alpha(\hat{k}_2)}{{\cal M}_k(\hat{k}_2)},\bar{\gamma}\geq\dfrac{\gamma(\hat{k}_2)}{{\cal M}_k(\hat{k}_2)}\right\},\label{eq:regionConst2}\\
\mathcal{R}_{{\sf mbr}}&=\left\{ (\bar{\alpha},\bar{\gamma}): \bar{\alpha}\geq\dfrac{\alpha_{{\sf mbr}}}{{\cal M}},\bar{\gamma}\geq\dfrac{\gamma_{{\sf mbr}}}{{\cal M}}\right\}, \,\,\,\,\textrm{and}\\
\mathcal{R}_{{\sf msr}}&=\left\{ (\bar{\alpha},\bar{\gamma}): \bar{\alpha}\geq\dfrac{\alpha_{{\sf msr}}}{{\cal M}},\bar{\gamma}\geq\dfrac{\gamma_{{\sf msr}}}{{\cal M}}\right\},
  \end{align}
where ${\cal R}_1$ is the tradeoff region corresponding to {\em Construction $1$}, and $\alpha(\hat{k}_1), \gamma(\hat{k}_1)$ and ${\cal M}(\hat{k}_1)$ in (\ref{eq:regionConst1}) are given by (\ref{eq:alpha-d=k}), (\ref{eq:gamma-d>k}), and (\ref{eq:capacity-d=k}), respectively, evaluated at $\hat{k}=\hat{k}_1$ and $\hat{n} = n+\hat{k}_1-k$.
Similarly, ${\cal R}_2$ corresponds to {\em Construction 2}, and $\alpha(\hat{k}_2), \gamma(\hat{k}_2)$ and ${\cal M}_k(\hat{k}_2)$ in (\ref{eq:regionConst2}) are given by (\ref{eq:alpha-d=k}), (\ref{eq:gamma-d=k}) and (\ref{eq:capacity-d>k}), respectively, evaluated at $\hat{k}=\hat{k}_2$ and $\hat{n} = n+\hat{k}_1-d$. The regions corresponding to the MBR and the MSR points are evaluated at the given $k$ and $d$.
\end{theorem}

\section{Conclusion}\label{sec:conclusion}
Determining the achievable region in the tradeoff between storage and exact repair bandwidth for distributed storage systems is an important problem that is still open in general, with this region fully characterized only for the $(4,3,3)$ case \cite{T13}. This paper makes a contribution towards solving this problem by proposing new  constructions of regenerating codes with exact repair that achieve new points in the tradeoff region leading to  improved inner bounds. Our code constructions are a generalization of the codes in \cite{E13} and \cite{SK13}. The main idea is to construct new codes using already known code constructions for systems with smaller parameters. Our constructions have two key ingredients: (i) Use as a building block a  minimum storage code (MSR) with optimal repair bandwidth for all possible repair degrees, and (ii) Allow a heterogeneous repair bandwidth that depends on the nodes contacted for repair. The code is then ``homogenized" by appending together all of its $n!$ permutations as described in \cite{ERHP13}. This method leads to new achievable points on the tradeoff and result in tighter inner bounds for all systems having more than one parity node ($n>k+1$).

\bibliographystyle{ieeetran}
\bibliography{IEEEabrv_sg,DSS}

\begin{thebibliography}{10}
\providecommand{\url}[1]{#1}
\csname url@samestyle\endcsname
\providecommand{\newblock}{\relax}
\providecommand{\bibinfo}[2]{#2}
\providecommand{\BIBentrySTDinterwordspacing}{\spaceskip=0pt\relax}
\providecommand{\BIBentryALTinterwordstretchfactor}{4}
\providecommand{\BIBentryALTinterwordspacing}{\spaceskip=\fontdimen2\font plus
\BIBentryALTinterwordstretchfactor\fontdimen3\font minus
  \fontdimen4\font\relax}
\providecommand{\BIBforeignlanguage}[2]{{%
\expandafter\ifx\csname l@#1\endcsname\relax
\typeout{** WARNING: IEEEtran.bst: No hyphenation pattern has been}%
\typeout{** loaded for the language `#1'. Using the pattern for}%
\typeout{** the default language instead.}%
\else
\language=\csname l@#1\endcsname
\fi
#2}}
\providecommand{\BIBdecl}{\relax}
\BIBdecl

\bibitem{DGWWR10}
A.~G. Dimakis, P.~G. Godfrey, Y.~Wu, M.~J. Wainwright, and K.~Ramchandran,
  ``Network {C}oding for {D}istributed {S}torage {S}ystems,'' in \emph{{IEEE}
  Transactions on Information Theory}, vol.~56, Sep. 2010, pp. 4539--4551.

\bibitem{RSK11}
K.~V. Rashmi, N.~B. Shah, and P.~V. Kumar, ``Optimal {E}xact-{R}egenerating
  {C}odes for {D}istributed {S}torage at the {MSR} and {MBR} {P}oints via a
  {P}roduct-{M}atrix {C}onstruction,'' in \emph{{IEEE} Transactions on
  Information Theory}, vol.~57, Aug. 2011, pp. 5227--5239.

\bibitem{PDC11}
D.~Papailiopoulos, A.~Dimakis, and V.~Cadambe, ``Repair {O}ptimal {E}rasure
  {C}odes through {H}adamard {D}esigns,'' in \emph{Proceedings of the 49th
  Annual Allerton Conference on Communication, Control, and Computing}, Sep.
  2011, pp. 1382--1389.

\bibitem{CJMRS13}
V.~Cadambe, S.~Jafar, H.~Maleki, K.~Ramchandran, and C.~Suh, ``Asymptotic
  {I}nterference {A}lignment for {O}ptimal {R}epair of {MDS} codes in
  {D}istributed {S}torage,'' in \emph{{IEEE} Transactions on Information
  Theory}, vol.~59, May 2013, pp. 2974--2987.

\bibitem{WTB11}
Z.~Wang, I.~Tamo, and J.~Bruck, ``On {C}odes for {O}ptimal {R}ebuilding
  {A}ccess,'' in \emph{Proceedings of the 49th Annual Allerton Conference on
  Communication, Control, and Computing}, 2011, pp. 1374--1381.

\bibitem{T13}
C.~Tian, ``Rate {R}egion of the $(4,3,3)$ {E}xact-{R}epair {R}egenerating
  {C}odes,'' in \emph{Proceedings of IEEE International Symposium on
  Information Theory (ISIT)}, Jul. 2013, pp. 1426--1430.

\bibitem{TAV13}
C.~Tian, V.~Aggarwal, and V.~A. Vaishampayan, ``Exact-{R}epair {R}egenerating
  {C}odes via {L}ayered {E}rasure {C}orrection and {B}lock {D}esigns,'' in
  \emph{Proceedings of IEEE International Symposium on Information Theory
  (ISIT)}, 2013, pp. 1431--1435.

\bibitem{SK13}
\BIBentryALTinterwordspacing
B.~Sasidharan and P.~V. Kumar, ``High-{R}ate {R}egenerating {C}odes {T}hrough
  {L}ayering,'' in \emph{arxiv.org}, Mar. 2013. [Online]. Available:
  \url{http://arxiv.org/abs/1301.6157}
\BIBentrySTDinterwordspacing

\bibitem{E13}
T.~Ernvall, ``Exact-regenerating codes between mbr and msr points,'' in
  \emph{arXiv:1304.5357v1}, 2013.

\bibitem{ERHP13}
T.~Ernvall, S.~{E}l Rouayheb, C.~Hollanti, and H.~V. Poor, ``Capacity and
  {S}ecurity of {H}eterogeneous {D}istributed {S}torage {S}ystems,'' in
  \emph{Proceedings of IEEE International Symposium on Information Theory
  (ISIT)}, 2013, pp. 1247--1251.

\bibitem{SR10}
C.~Suh and K.~Ramchandran, ``Exact-{R}epair {MDS} {C}odes for {D}istributed
  {S}torage using {I}nterference {A}lignment,'' in \emph{Proceedings of IEEE
  International Symposium on Information Theory (ISIT)}, Jun. 2010, pp.
  161--165.

\bibitem{ATVC13}
\BIBentryALTinterwordspacing
V.~Aggarwal, C.~Tian, V.~A. Vaishampayan, and Y.~R. Chen, ``Distributed {D}ata
  {S}torage {S}ystems with {O}pportunistic {R}epair,'' in \emph{arxiv.org},
  Nov. 2013. [Online]. Available: \url{http://arxiv.org/abs/1311.4096}
\BIBentrySTDinterwordspacing

\end{thebibliography}

\end{document}